\documentclass[aps,nofootinbib,11pt]{revtex4}
\usepackage{amsfonts,amssymb,amsmath,amsthm,graphicx}
\newcommand{\nc}{\newcommand}
\nc{\al}{\alpha}
\nc{\ga}{\gamma}
\nc{\de}{\delta}
\nc{\ep}{\epsilon}
\nc{\ze}{\zeta}
\nc{\et}{\eta}
\renewcommand{\th}{\theta}
\nc{\ka}{\kappa}
\nc{\la}{\lambda}
\nc{\rh}{\rho}
\nc{\si}{\sigma}
\nc{\ta}{\tau}
\nc{\up}{\upsilon}
\nc{\ph}{\phi}
\nc{\ch}{\chi}
\nc{\ps}{\psi}
\nc{\om}{\omega}
\nc{\Ga}{\Gamma}
\nc{\De}{\Delta}
\nc{\La}{\Lambda}
\nc{\Si}{\Sigma}
\nc{\Up}{\Upsilon}
\nc{\Ph}{\Phi}
\nc{\Ps}{\Psi}
\nc{\Om}{\Omega}
\nc{\ptl}{\partial}
\nc{\del}{\nabla}
\nc{\be}{\begin{equation}}
\nc{\ee}{\end{equation}}
\nc{\ba}{\begin{eqnarray}}
\nc{\ea}{\end{eqnarray}}
\nc{\bi}{\bibitem}

\newcommand{\beq}{\begin{equation}}
\newcommand{\eeq}{\end{equation}}
\newcommand{\beqn}{\begin{eqnarray}}
\newcommand{\eeqn}{\end{eqnarray}}

\begin{document}
\title{\Large 
  Weyl corrections to holographic conductivity 
}
\author{Adam Ritz and John Ward}
\affiliation{
  Department of Physics and Astronomy, University of Victoria,
  Victoria, BC, V8P 5C2, Canada
  }
\date{November 2008}
\begin{abstract}
\noindent
For conformal field theories which admit a dual gravitational description in anti-de Sitter space, electrical transport properties, such
as conductivity and charge diffusion, are determined by the dynamics of a U(1) gauge field in the bulk and thus obey universality relations
at the classical level due to the uniqueness of the Maxwell action. We analyze corrections to these transport parameters due to 
higher-dimension operators in the bulk action, beyond the leading Maxwell term, of which the most significant involves
a coupling to the bulk Weyl tensor. We show that the ensuing corrections to conductivity and the diffusion constant 
break the universal relation with the U(1) central charge observed at  leading order, but are nonetheless subject to interesting bounds
associated with causality in the boundary CFT.
\end{abstract}
\maketitle


\section{Introduction and Summary}
\label{sec:cc}
\noindent

For a conformal field theory (CFT) in a thermal ensemble, the fact that the temperature $T$ is the only scale naturally implies the presence of
 two characteristic regimes distinguished by whether the length scale being probed is large or small relative to $1/T$. 
At short distances, temperature is essentially irrelevant
and the theory is characterized by various central charges which dictate the leading singular behaviour of the
correlation functions of conserved currents. In contrast, at long distances the temperature becomes very important and the theory is best 
described by thermodynamic parameters and transport coefficients. Despite the scaling symmetry of 
a CFT, characterizations of these regimes are generally not related in spacetime dimensions $d>2$ \cite{sachdev}. However for CFTs
which exhibit a dual description via classical gravity in Anti-de Sitter (AdS) space \cite{magoo}, it turns out that all these defining parameters
of the theory at different scales are indeed interdependent \cite{Kovtun:2008kx}. 

Focusing on a conserved U(1) `electric' current $J_\mu$, recall that for a zero temperature CFT the Euclidean correlator
$\langle J_\mu(x) J_\nu (0)\rangle$ is determined uniquely by a U(1) central charge $k$. For $T\neq 0$ the equilibrium
state is characterized in turn by the charge susceptibility $\ch$, and since $T$ is the only scale we can write
$\ch=k' T^{d-2}$ in terms of another dimensionless constant $k'$. Furthermore one can also consider dynamical
transport coefficients associated with $J_\mu$, such as the dc conductivity $\si$.
In \cite{Kovtun:2008kx} it was shown that for CFTs which have AdS duals, and in the classical limit
where the bulk action reduces to Einstein-Maxwell theory, all of these quantities are in fact related as follows
\be
 \si =  \frac{\ch}{4\pi T} \frac{d}{d-2} = \left[\frac{1}{8\pi^{d/2+1}} \frac{d}{d-2}\left(\frac{4\pi}{d}\right)^{d-2} \frac{\Ga(d/2)^3}{\Ga(d)}\right]  k T^{d-3}. \label{cond}
\ee
This story has parallels with a similar relationship between the central charge $c$ of the CFT,  as determined by the energy-momentum tensor
 two-point function, and the entropy and shear viscosity which are in turn related to it in CFTs with classical AdS duals \cite{kss,bl,Kovtun:2008kw},
 \be
  \eta = \frac{s}{4\pi} = \left[\frac{1}{16\pi^{d/2+1}} \frac{d-1}{d+1}\left(\frac{4\pi}{d}\right)^{d} \frac{\Ga(d/2)^3}{\Ga(d)}\right] c T^{d-1}.
   \ee
 This link also motivated the conjecture that in certain classes of systems (excluding at least those with a non-zero chemical potential for the 
 U(1) charge), the relation for the conductivity in (\ref{cond}) might actually be a lower bound saturated by relativistic CFTs with classical AdS duals.
 
 An immediate question that arises, is how this picture is modified as one goes beyond the classical AdS/CFT limit and considers
 various higher-derivative corrections in the bulk that will necessarily arise through quantum effects of various kinds. In the 
 example of ${\cal N}=4$ SYM, this means going beyond the large-$N$ limit and/or including finite 't Hooft coupling
 corrections. Indeed, while relations such as (\ref{cond}) appear quite nontrivial for the CFT, they arise almost trivially from
 the bulk perspective since all these quantities are determined by the normalization of the Maxwell action. Thus the relationship
 in (\ref{cond}) is ensured by the uniqueness of the minimal dimension gauge invariant operator for a U(1) vector field. This uniqueness is clearly
 broken on including higher-derivative corrections, and so we can anticipate the simple interdependence in (\ref{cond}) to be modified. 
 However, it is interesting to see how these corrections arise, and the form of any associated  constraints. A similar analysis was
 recently carried out for shear viscosity by looking at curvature-squared terms in the gravitational Lagrangian \cite{myers,kp,bm}. In many respects the 
 present problem is simpler, as one can show that the background uncharged black hole geometry remains a solution to
 all orders. The dynamics of a U(1) gauge field $A_\mu$ in this background decouples from any perturbations in the metric, and thus 
 the problem reduces to analyzing the quasinormal modes for $A_\mu$ with the  generalized dynamics, 
 \be
  {\cal L} =\frac{1}{4g_{d+1}^2} X^{\mu\nu\rh\si} F_{\mu\nu} F_{\rh\si} = \frac{1}{4g_{d+1}^2} 
  \left( F^{\mu\nu}F_{\mu\nu} - 4\gamma C^{\mu\nu\rh\si}F_{\mu\nu} F_{\rh\si} + \cdots\right),  \label{Lnext}
 \ee
 where, since the background geometry will be an Einstein metric, we will argue that there is a unique tensorial 
 structure correcting the Maxwell term at leading order in derivatives, arising from a coupling to the Weyl tensor and
 leading to the dimension-six operator (given $[g^2_{d+1}]=3-d$) in (\ref{Lnext}) parametrized by the constant $\gamma$.
 Other curvature couplings simply provide constant shifts of  $g^2$ when considering linearized gauge field fluctuations about 
 the background. 
  
 This setting provides a very clean test of the universality relation (\ref{cond}) at higher order,  because the geometry dual to the 
 zero temperature state, namely pure AdS, is Weyl flat. It follows immediately that the  2-point function $\langle J_\mu(x) J_\nu (0)\rangle$
 at $T=0$, and thus the central charge $k$ \cite{freedman,Kovtun:2008kx}, are uncorrected by turning on the perturbation $\gamma$,
 \be
  k(\gamma) = k(\gamma=0).
 \ee
  Any finite correction to
 the diffusion constant $D=\si/\ch$ will then reflect corrections to $\si$ and/or $\ch$ which are not dictated solely by the central charge. 
 Following the Minkowskian AdS/CFT machinery \cite{Son:2007vk}, and also the membrane paradigm prescription, which both give consistent results
 we compute the corresponding corrections to conductivity and the  diffusion constant, 
 obtaining the results (for $d=4$),
\be
 \si = \frac{\pi L T}{g_5^2} \left(1+\frac{8 \gamma}{L^2}\right), \;\;\;\;\;\;\;\; D= \frac{\si}{\ch} =\frac{1}{2\pi T}\left(1+\frac{16 \gamma}{L^2}+\cdots \right),
  \ee
 where $L$ is the curvature scale, and the result for $D$ is perturbative in $\gamma$. 
 Thus we indeed find that the universality relation fails to hold at higher order. While such corrections
 are to be expected on general grounds, our primary aim was to explore any patterns in how they arise and indeed to see if there
 are any generic constraints. The fact that there is only one independent tensor structure at this order is already a significant
 simplification.  Computing corrections for $d=3$ and $d=6$, we observe similar results with the corrections to $D$ and $\si$ all
 having the same sign as for $d=4$. Given these generic results obtained within effective field theory, it is interesting to explore
 explicit examples (which in the case of $\eta/s$ are often more restrictive \cite{R4}), and for the Weyl coupling $\gamma$ we 
 note the following:
 
\begin{itemize}
\item {\it causality constraints}: Although the possibility of IR effective field theory manifestations of UV causality constraints  has had some
attention recently \cite{aadnr}, this issue is rather subtle in curved space. Indeed, as reviewed below, QED in curved space does lead to
Weyl couplings at 1-loop \cite{dh} in a form which do apparently allow for superluminal propagation of certain photon polarizations; however,
this IR effect in curved space does not actually represent a violation of causality. Nonetheless, if we go beyond effective field theory 
and treat the Weyl-corrected action at the classical level as it stands, then the AdS/CFT context provides an interesting arena to review these issues as the
boundary causal structure is fixed and thus superluminal propagation in the CFT should indeed reflect a violation of causality. Following
the argument of Brigante et al.  \cite{myers}, we observe that a lower bound can be placed on $\gamma$, namely $\gamma > -L^2/16$, to avoid the
possibility of superluminal transport by metastable quasi-particles in the CFT (we also observe that an upper bound on $\gamma$  seems to 
be required to avoid modes becoming ghost-like near the horizon). This conclusion is analogous to the result of \cite{myers} for
corrections to shear viscosity, and leads to the constraints (for $d=4$),
\be
 \si(\gamma) > \frac{1}{2} \si(\gamma=0),\;\;\;\;\;\; D(\gamma) = \frac{\si}{\ch} > D(\gamma=0)\times 0.3617...,  \label{limits}
\ee  
which, while not directly supporting the conjectured bound in \cite{Kovtun:2008kx}, does suggest that 
it cannot be violated by orders of magnitude. In this regard, the story has parallels with the analysis of curvature-squared
 corrections to $\eta/s$  \cite{myers,kp,bm}.\footnote{A possible counter-example for $\eta/s > 1/(4\pi)$ has been
discussed in \cite{kp}, corresponding to {\cal N}=2 SYM with SO/Sp gauge groups, where the curvature-squared correction can be linked to
the difference of the two central charges $a-c$ at ${\cal O}(N)$ \cite{bng}, although even in this case the full background reproducing both central charges
at ${\cal O}(N)$ is not known.}

\item {\it quantum corrections}: In any background in which additional charged matter fields are integrated out below their mass threshold,
the Weyl coupling $C^{\mu\nu\rh\si}F_{\mu\nu} F_{\rh\si}$ is generated at 1-loop, with a coefficient $\gamma \sim \al/m^2$ 
first computed (for $d+1=4$) by Drummond and Hathrell \cite{dh}. To read off the results, we have first to take into account the threshold corrections
to the U(1) gauge coupling, which arise from curvature couplings to $R$ and $R_{\mu\nu}$. Working with the resulting low-energy
gauge coupling, the (renormalized) expression for $\gamma$ takes the form,
\be
 \gamma^{d=3}_{\rm 1-loop} = -\frac{\al (n_s+4n_f)}{1440 \pi m^2} \left(1+ {\cal O}\left(\frac{1}{(mL)^2}\right)\right) \label{1loop}
\ee
for $n_s$ complex scalars \cite{hs} and $n_f$ Dirac fermions \cite{dh} with generic mass $m$ and 
coupling $\al$ to the U(1) gauge field (neglecting logarithmic running of the gauge coupling above the threshold). 
We require $m\gg 1/L$, so the contribution to $\gamma$ is negative but intriguingly still well within 
the $(d=3)$ variant of the conjectured lower bound on $\gamma$ discussed above, and thus does not imply
superluminal propagation about the AdS black hole geometry. Note that  the requirement $m\gg 1/L$ means,
from the AdS/CFT perspective, a parametrically large bulk cutoff $\sim f(N,g^2N)/L^2$ scaling with $N$ or the 't Hooft coupling.

\item {\it $\al'$ corrections}: Beyond bulk quantum effects, it would be interesting to know if such corrections do arise at tree-level
within the ${\cal O}(\al')$ expansion. We are not aware of any concrete compactifications which realize these Weyl couplings, but
within  ${\cal N}$=4 SYM with $\gamma \sim {\cal O}(\al')$ the higher order contribution to $D$ would be of ${\cal O}(1/\sqrt{g^2N})$ reflecting a 
non-universal correction away from the large 't Hooft coupling limit. 
\end{itemize} 

Having summarized the results here, in the next section we discuss the general constraints on higher-derivative corrections, motivating
(\ref{Lnext}) as the leading irrelevant operator correction. In Section~3, we perform the computations of $\si$ and $D$ for various backgrounds using 
both the conventional AdS/CFT prescription for linear response and also a variant of the membrane paradigm. We finish with a discussion 
in Section~4, focusing in particular on possible causality constraints on the parameter $\gamma$.

\section{General Current Sources}

Working within the framework of linear response, we will consider the transport properties associated with a conserved current in
an uncharged thermal state. This means that the dual gravitational background should be an uncharged black brane, and so we 
can write the action for the bulk U(1) gauge field $A_\mu$, which at the boundary is the source for the current, quite generally
as,
\beq
 S =  \int d^{d+1}x \sqrt{-g}\left( -\frac{1}{4g_{d+1}^2}X^{\mu\nu\rh\si} F_{\mu\nu} F_{\rh\si} + {\cal O}(A^3)\right),
\eeq
in terms of a tensor $X$ satisfying,
\beq
 X^{\mu\nu\rh\si} = X^{[\mu\nu][\rh\si]} = X^{\rh\si\mu\nu},
\eeq
which depends on the background metric. We have neglected terms of higher than quadratic order
since they will not contribute to the linearized fluctuation equations for an uncharged background which will necessarily have
$F^{(0)}_{\mu\nu}=0$. 

The relevant equations of motion take the form,
\ba
 && G_{\mu\nu} = \La g_{\mu\nu} + T^{A}_{\mu\nu\rh\si\gamma\de} F^{\rh\si} F^{\gamma\de}, \nonumber \\
 && \ptl_\mu \left( \sqrt{-g} X^{\mu\nu\rh\si} F_{\rh\si}\right) =0,
\ea
where $G_{\mu\nu}$ is the Einstein tensor, $\La=-d(d-1)/L^2$ is the cosmological constant, 
and $T^{A}$, which depends on $X$, determines the energy momentum tensor for $A_\mu$. We observe that
for any choice of $X^{\mu\nu\rh\si}$, a conventional uncharged black brane metric  $g^{(0)}_{\mu\nu}$ with $F^{(0)}_{\mu\nu}=0$ is a solution 
which we will take to describe the background. Perturbing about this background to linear order, the equations for $\de A_{\mu}$ and
the vector perturbation in the metric decouple, and so we can focus on Maxwell's equation in the unperturbed background,
\be
 \ptl_\mu \left( \sqrt{-g_{(0)}} X^{\mu\nu\rh\si}_{(0)} F_{\rh\si}(\de A)\right) = 0.
 \ee
 
 At this point, we see that treating linear perturbations about the uncharged background is a significant technical simplification, as
 we can study the general tensor structure $X$ in the uncorrected background, so that $g^{(0)}_{\mu\nu}$ is an Einstein metric.
 In general there are then only two independent geometrical tensor structures,
 \be
  X_{\mu\nu\rh\si} = \frac{a(g)}{2} g_{\mu[\rh}g_{\si]\nu} - b(g) C_{\mu\nu\rh\si},
 \ee 
 where $C_{\mu\nu\rh\si}$ is the Weyl tensor, and $a$ and $b$ are, within an effective field theory expansion, polynomial functions of 
 the metric and derivatives. 
The remaining tensor structures that we could have in general, $R_{\mu\nu}$ and 
 $R_{\mu\nu\rh\si}$ are reducible to this set for an Einstein metric. The leading order term in a derivative expansion, the
 Maxwell term, then corresponds to setting $a=1$ and $b=0$.
 \be 
  \left. X_{\mu\nu\rh\si}\right|_{\rm LO} = \frac{1}{2} g_{\mu[\rh}g_{\si]\nu}. \label{maxwell}
  \ee
  If we do not impose parity as a symmetry, then in 4D (i.e. $d=3$) we could also include a topological contribution proportional to $\ep_{\mu\nu\rh\si}$, 
  and in odd dimensions, one can have Chern-Simons terms. We will ignore these parity-odd terms in this paper.

 The leading-order corrections correspond to operators of dimension six (given $[g^2_{d+1}]=3-d$) and, due to the symmetries of the background, there
 are only two classes of terms. The first comprises pure derivative corrections to (\ref{maxwell}), e.g. operators like $F \Box F$ which,
 using various identities, can all be reduced to operators which are zero according to the background equations of motion,
 i.e. $(\nabla_\mu F^{\mu\nu})^2$ \cite{DvN}. These operators can only contribute at higher order and will be ignored here. The second
 class of dimension six terms are couplings to the curvature tensors and as discussed above for an Einstein metric, up to a constant 
 ``renormalization'' of the gauge coupling $g^2\rightarrow g^2_{\rm eff}$ that we will implicitly absorb, only the Weyl coupling provides an 
 independent structure. Thus we are led to consider a unique dimension-six operator as the leading correction to the equations of motion
 for linearized gauge field fluctuations,
 \be
  X_{\mu\nu\rh\si} = \frac{1}{2} g_{\mu[\rh}g_{\si]\nu} - 4\gamma C_{\mu\nu\rh\si},
 \ee 
parametrized by the (dimensionful)  constant $\gamma$.

\section{Corrections to conductivity and diffusion}

Following the arguments above, we can now limit our attention to the generalized Maxwell equation,
\be
 \ptl_\mu \left[ \sqrt{-g} \left( F^{\mu\nu} - 4 \ga C^{\mu\nu\rh\si} F_{\rh\si}\right)\right] = 0,
 \ee
 in terms of the background geometry dual to the thermal state. The relevant part of the geometry is the non-extremal AdS$_{d+1}$
 metric,
 \beq
ds_{d+1}^2 = \frac{r^2}{L^2}\left(-f(r)dt^2 + d\vec{x}^2 \right) + \frac{L^2}{r^2}\frac{dr^2}{f(r)}, \label{adsbh}
\eeq
where $f(r) = 1 - (r_0/r)^d$ in terms of the horizon radius $r_0$, or equivalently the Hawking temperature $T=r_0 d/(4\pi L^2)$.  
We will work primarily with $d=4$ in this section, for which the Weyl tensor has the following non-zero components,
\beqn
C_{0i0j} &=& \frac{f(r) r_0^4 \delta_{ij}}{L^6}, \hspace{1cm} C_{0r0r} = -\frac{3 r_0^4}{L^2 r^4}  \\
C_{irjr} &=& -\frac{\delta_{ij}r_0^4}{L^2r^4 f(r)}, \hspace{0.7cm} C_{ijkl} = \frac{r_0^4}{L^6}\delta_{ik}\delta_{jl}. \nonumber 
\eeqn
Note that two of these terms vanish on the boundary, whilst the $0i0j$ component vanishes on the horizon. As a point of interest,
these are the only nonzero components of the Weyl tensor for any five-dimensional `black' metric, for which $g_{rr}$ and $g_{00}$
are inversely related.

\subsection{Diffusion and conductivity within the membrane paradigm}

The apparent analogy between the AdS/CFT description of black hole geometries at the level of linear response, and the 
 membrane paradigm \cite{membrane,pw} has been noticed by many authors \cite{kss1,il,mst}. In the present context, working on the stretched horizon
at $r_{\ep}=r_0+\ep$, a natural `membrane current' to define is the momentum conjugate to $A_\mu$ for a radial foliation,
\be
 j^{\mu} = \left . \sqrt{-g} X^{\mu r\rh\si} F_{\rh\si}\right|_{r_{\ep}},
\ee
which is necessarily conserved $\ptl_\mu j^\mu$ on account of the equations of motion. We should note that this differs 
slightly from the original membrane current \cite{pw},
\be
 J_m^\mu = \left. n_\nu X^{\mu\nu\rh\si} F_{\rh \si}\right|_{r_\ep},
\ee
by a factor of the induced metric on the stretched horizon for which $n_\nu$ is a unit  radial normal vector. The distinction
between these two definitions has recently been noted in \cite{il}, and while we will focus on the former definition, we
will also comment on the conductivity for $J_m$.

Within the hydrodynamic regime, $j^0$ evolves according to the diffusion equation (Fick's Law),
\be
 \ptl_0 j^0 = D \nabla^2 j^0,
\ee
implying the presence of a mode with the diffusive dispersion relation $\om=-iD q^2$ in terms of the
diffusion constant $D$. If we apply this picture at the stretched horizon,  generalizing the Maxwell case \cite{kss1}, we find
upon solving the equations in radial gauge for $A_\mu$ -- assumed to be slowly varying in directions tangent to 
the stretched horizon -- that 
\be
 \left. F_{rx}\right|_{r_\ep} = \left.\sqrt{-\frac{X^{x0x0}}{X^{r0r0}} }F_{0x}\right|_{r_\ep},
\ee
where we have singled out the spatial $x$-coordinate. Using this relation, one can write down Ohm's Law on the
stretched horizon, $j^x = \si E^x$, with the conductivity given by
\be
 \si = \left .2\sqrt{-g} \sqrt{-X^{x0x0}X^{rxrx}}\right|_{r_\ep} = \frac{1}{g_5^2} \frac{r_0}{L} \left(1+\frac{8 \gamma}{L^2}\right). \label{condm}
 \ee
 In this expression, we have rescaled to unity the charge $e$ that arises by weakly gauging the global U(1) symmetry, which
 we will do throughout the paper. The physical conductivity is then $\si \rightarrow \si e^2$.

Moreover, by solving the equation for $A_0$ near the horizon following \cite{kss1,myers}, we obtain Fick's law in the form 
$j^x = - D \ptl_x j^0$, with the diffusion constant
\be
 D = -\sqrt{-g} \left. \sqrt{-X^{x0x0} X^{xrxr}}\right|_{r_\ep} \int_{r_\ep}^{\infty} \frac{dr}{\sqrt{-g} X^{0r0r}}. \label{Dm}
\ee
Note that this expression naturally factorizes in accord with the Einstein relation $D = \si/\ch$, and so with the 
conductivity given by (\ref{condm}) we can also read off the susceptibility $\ch$ from (\ref{Dm}).
Evaluating the components of $X$ in terms of the metric and the Weyl tensor, we find that provided $0< \gamma < L^2/24$ (and with
a suitable analytic continuation outside this range)
the diffusion constant takes the form
\be
 D = \frac{1}{2\pi T}\left(\frac{1+8\gamma/L^2}{4\sqrt{6\gamma/ L^2}}\ln \frac{L+2\sqrt{6\gamma}}{L-2\sqrt{6\gamma}}\right) \sim 
     \frac{1}{2\pi T} \left( 1 + \frac{16 \gamma}{L^2}+\cdots \right).
 \ee
We will verify this result, treating $\gamma$ perturbatively, by explicit computations following the AdS/CFT
prescription in a later subsection. Indeed, the correspondence between the AdS/CFT and membrane prescriptions for 
computing transport coefficients has recently been put on a firmer footing \cite{il}, and the current setting extends this equivalence 
beyond the examples considered in \cite{il}.

More generally, we can consider the diffusion constant in the ($d+1$)-dimensional non-extremal AdS background (\ref{adsbh}). After computing 
the various components of the Weyl tensor, bearing in mind that it is identically zero in two-dimensions, we find the following general result,
\be
\label{eq:generalD}
D = \frac{d}{(d-2)} \frac{1}{4\pi T} \left(1+\frac{2 d(d-2)\gamma}{L^2}+\cdots  \right).
\ee

For comparison, the conductivity associated with the membrane current, $J_m^x = \si_m \hat{E}^x$, where $\hat{E}^x$ is
the electric field measured in a local orthonormal  frame at the stretched horizon \cite{il}, is 
\be
 \si_m = \frac{1}{g_5^2} \left(1+\frac{8 \gamma}{L^2}\right),
 \ee
 which differs from the more conventional  temperature dependence in Eq.~(\ref{cond}) for $d\neq 3$.

\subsection{Corrections to the diffusion constant within AdS/CFT}
Using the conventional AdS/CFT prescription for linear response theory, we can also extract the relevant parameters directly from the retarded propagators.
Fixing $d=4$, it is convenient to employ the radial gauge $A_r = 0$, and also to work in the Fourier-space representation of the gauge field,
\beq
A_i(t,z,r) = \int \frac{d^4 q}{(2 \pi)^4} e^{-i\omega t + i q z}A_i(\om,q,r),
\eeq
where we single out the $z$ coordinate for convenience. 
The computation is also more tractable if we re-write the metric in terms of its Hawking temperature using a new set of coordinates \cite{Policastro:2002se},
\beq
ds^2_5 = \frac{\alpha^2 L^2}{u}\left(-f(u)dt^2+d\vec{x}^2 \right)+ \frac{L^2}{4u^2f(u)}du^2,
\eeq
where
\beq
\alpha = \pi T_H, \hspace{1cm} f(u)=1-u^2.
\eeq
Using this metric,  the non-zero components of the Weyl tensor become:
\beqn
C_{0i0j} &=& \delta_{ij}f(u)\alpha^4 L^2,\hspace{1cm} C_{0u0u} = - \frac{3 \alpha^2 L^2}{4u},\\
C_{iuju} &=& - \frac{\delta_{ij} \alpha^2 L^2}{4 u f(u)}, \hspace{1.5cm}  C_{ijkl} = \delta_{ik} \delta_{jl} \alpha^4 L^2.\nonumber 
\eeqn
We must now solve the relevant component expansion of the modified Maxwell equations. After some algebra we find the following expressions in component
form:
\beqn
0 &=& (1+2 Q)(\omega q A_z + q^2 A_0)-4 \alpha^2 f u A_0''(1-6Q)+48Qf \alpha^2A_0'  \label{eom} \\
0&=& \omega A_0'(1-6Q)+fqA_z'(1+2Q) \nonumber \\
0&=& A_z''(1+2Q) + \frac{(\omega^2 A_z+\omega q A_0)}{4 \alpha^2 f u}(1+2Q) + A_z' \frac{f'}{f}\left(1+2Q+4Q\frac{f}{uf'} \right) \nonumber \\
0&=& (1+2Q)\left(\frac{\omega^2}{2fu}A_{\beta}+2\alpha^2 f A_{\beta}^{\prime \prime} \right)-\frac{q^2(1-2Q)}{2u}A_{\beta} 
+ 2 \alpha^2 f' A_{\beta}' \left(1+2Q(1+\frac{2f}{uf'})\right), \nonumber
\eeqn
where we have defined $Q = 4\gamma u^2/L^2$
and the sub-script $\beta$  runs over the $x,y$ directions (since we have singled out $z$ in the definition of the Fourier transform). 

We can decouple the first two equations in (\ref{eom}) by solving the first for $A_z$ and substituting into the second, which then takes
the symbolic form $A_0'''+\alpha_2 A_0'' + \alpha_1 A_0' = 0$ with coefficients $\al_1$ and $\al_2$ which can be read off from (\ref{eom}).
In order to solve this equation we must consider the behaviour of $A_0'$ at the horizon, where the solution is singular. However, there is 
also an additional singular point present when $6Q=1$, i.e. $u^2=L^2/(24\gamma)$, which we will remove with the constraint
\beq
\gamma < \frac{L^2}{24},
\eeq
to be interpreted in more detail in the final section. As in the standard case, the horizon at $u=1$ is a singular point for the 
differential equation, and imposing causal incoming boundary conditions there, the solution is required to take the form
$A_0' = (1-u)^{-i\om/(4\al)}F(u)$ where $F(u)$ is regular. This singular behaviour is unchanged from the Maxwell case
with $\gamma =0$ \cite{Policastro:2002se}.

Since we want to consider the large wavelength limit where both $\omega$ and $q^2$ are small, we will use a combined
perturbative expansion for $F$ in $\om$, $q^2$ and $\gamma$, 
\beq
F(u) \sim F_0 + \omega F_1 + q^2 G_1 + \gamma H_1 + \gamma \omega H_2 + \gamma q^2 H_3 + \cdots
\eeq
Expanding the equations to the appropriate order, we find that the perturbative solution takes the form,
\be
 F(u) = F_0 \left[ 1 + 24\gamma \frac{u^2}{L^2} + \frac{i\om}{4\al}\ln\left(\frac{2u^2}{1+u}\right) + \frac{q^2}{4\al^2}\ln\left(\frac{1+u}{2u}\right)\right] 
  + \gamma \omega H_2 + \gamma q^2 H_3 + \cdots
  \ee
where for completeness, the two higher-order contributions are given by
\beqn
H_2 &=& \frac{2 iF_0}{\alpha L^2} \left(3u^2 \ln \left(\frac{2u^2}{1+u} \right) - 2\ln u +3u^2\ln(2) \right) \nonumber \\
H_3 &=& - \frac{2 F_0}{\alpha^2 L^2} \left( 3 u^2 \ln(\frac{1+u}{2u})+2u + \ln u - 2\ln(1+u)\right).
\eeqn
The constants of integration have been fixed by requiring regularity at the horizon. In fact we have normalised the solutions so
that they vanish at the horizon, with the exception of $H_1$ which should remain finite in order to obtain regular 
solutions for $H_2, H_3$.

Given this solution for $A_0'$, the corresponding solution for $A_z$ is determined to leading order in $\gamma$ as follows,
\beq
A_z = \frac{4 \alpha^2 f u}{\omega q}A_0'' \left(1-\frac{32\gamma u^2}{L^2} \right)-\frac{192 \gamma u^2}{L^2} \frac{f \alpha^2}{\omega q}A_0' - \frac{q}{\omega}A_0,
\eeq
which, upon defining the boundary sources $A_t^0=A_t(u\rightarrow 0)$, $A_z^0=A_z(u\rightarrow 0)$, allows us to fix  $F_0$,
\be
 F_0 = \frac{A^0_z \omega q + q^2 A^0_t}{2i \al \omega(1-8\gamma/L^2) - (1+8\gamma/L^2) q^2}.
 \ee
Following the Minkowskian AdS/CFT prescription \cite{Son:2002sd}, this is enough information to extract the 
retarded correlator $G_{tt}$ for the charge density $j^0$. In particular, the solution for $A_0'$ reduces near the boundary to $A_0' \sim F_0 +{\cal O}(\om,q^2)$,
while the bulk action for $A_0$ is given by
\be
S = \int d^5 x \frac{\alpha^2 L}{g_5^2}\left(1-\frac{24 \gamma u^2}{L^2} \right) A_0'^2  + \cdots
\ee
It follows that the retarded correlator takes the form,
\be
 G_{tt} = \frac{\chi D q^2}{(i\omega-Dq^2)} = \frac{2 \al^2 L q^2}{g_5^2}\frac{1}{(2i\al \omega(1-8\gamma/L^2) - (1+8\gamma/L^2) q^2)},
  \ee
  where, making use of the Einstein relation $\si = \ch D$, we can read off the dc conductivity
  \be
  \si =  \frac{\alpha L}{g_5^2} \left(1+\frac{8 \gamma}{L^2} \right) +\cdots, \label{eq:cond}
  \ee
  and the diffusion constant,
  \be
   D =  \frac{\si}{\ch}  = \frac{1}{2 \alpha}\left(1+\frac{16 \gamma}{L^2} \right)+\cdots,
     \ee
     which we note are in agreement with the known results for $\gamma=0$, and our earlier computations using the membrane current.

\subsection{Corrections to conductivity within AdS/CFT}
We can also verify the calculation of the conductivity (and thus the Einstein relation) more directly from the 
 spatial correlator $G_{xx}$, by solving the gauge field equations of motion for $A_x$, which corresponds to the 4th equation in (\ref{eom}). 
The equation again has a singular point at the horizon $u=1$, and requiring an ingoing boundary condition as above, we have
$A_{x} = (1-u)^{-i\om/(4\al)}G(u)$ with $G(u)$ a regular function.
This again gives us an equation of the schematic form $G'' + A G' + B G =0$.
Although the conductivity only requires knowledge of $G_{xx}(\om,q^2=0)$, for completeness we will look for a full perturbative solution in $\om$ and
$q^2$ of the form 
\beq
G(u) = G_0 +  \omega G_1  +  q^2 H_1+ \gamma J_1  + \gamma \omega J_2 + \cdots
\eeq
The regularized solution is given by,
\ba
 G(u) &=& G_0\left[ 1+\frac{i\om}{4\al} \ln(1+u)+\frac{q^2}{8\al^2}\left( {\rm Li}_2(u)+{\rm Li}_2(1+u)+\ln(u)\ln(1+u)\right)\right]  \nonumber\\
   && \;\;\;\; + (\om +q^2) A + \gamma J_1  + \frac{\gamma \om}{2\al L^2} \left( 8i G_0 u + i J_1\ln(1+u)L^2 + 4 B \al L^2\right) + \cdots 
\ea
where $A$ and $B$ are integration constants that drop out once we express $A_x$ in terms of the source 
$A^0_x = A_x (u\rightarrow 0)$, i.e. $G_0  = A^0_x - A \omega - \gamma J_1 - 2\gamma \omega B$.
Given the normalization of the on-shell action for $A_x$, 
\beq
S = -\int d^5 x \frac{\alpha^2 L f}{ g_5^2}\left(1+\frac{8 \gamma u^2}{L^2} \right) A_x'^2+ \cdots,
\eeq
which fixes the induced coupling to the boundary current $J_x \propto A_x'(u\rightarrow 0)$, 
the conductivity can be obtained in one of two ways . Expanding the solution for $A_x$ near the boundary 
$A_x(u) \sim A_x^0 + g_5^2/(2\al^2 L) u J_x + \cdots$, determines the current $J_x$ and 
the electric field $\ptl_t A_x^0$ in the dual field theory, and thus from Ohm's Law we can read off the conductivity $\si = {\rm Re}[J_x/(i\om A_x^0)]$.
Alternatively,  we obtain the correlator $G_{xx}$ from the action in analogy with the earlier treatment of $G_{tt}$, and 
the conductivity (with $e^2=1$ as above)  is then given by the Kubo formula $\si(\om) =  -{\rm Im} ( G_{xx}(\omega, 0)/\om)$. In either
case we obtain
\beq
\sigma =  \frac{\alpha L}{g_5^2}\left(1 + \frac{8 \gamma }{L^2} \right) + \cdots
\eeq
in agreement with the result extracted from the $tt$ correlator (\ref{eq:cond}).


\subsection{Corrections to conductivity in 3 dimensions}
We can of course also consider what happens in other gravity duals, taking for example the background (\ref{adsbh}) with $d=3$, as would
arise from the near horizon limit of a stack of  black M2-branes after dimensionally reducing over the transverse space.
Since the bulk is now 4-dimensional, we could also add a parity-odd topological term $\th F \tilde{F}$ to the action. It is known that this
results in a contribution to the  Hall current, i.e. $\si_{ij} = \si_H \ep_{ij}$, but this case has been covered in the literature \cite{hall,il} so we
not pursue it further here. Retaining the Maxwell term and the Weyl correction, we note that the non-zero components 
of the four-dimensional Weyl tensor are given by:
\beqn
C_{0i0j} &=& \frac{2 f \delta_{ij} L^2 \alpha^4}{u} \hspace{1cm} C_{0u0u} = -\frac{\alpha^2 L^2}{u} \nonumber \\
C_{iuju} &=& - \frac{\alpha^2 L^2 \delta_{ij}}{2 u f} \hspace{1.3cm} C_{ijkl} = \frac{4 \alpha^4 L^2}{u} \delta_{ik} \delta_{jl},
\eeqn
where $u=r_0/r$ in this case, while $\al=2\pi T/3$ and $f(u)=1-u^3$.
It turns out to be difficult to compute the $tt$ correlator in this theory even perturbatively. However we can compute the
spatial $yy$ component, assuming that we align the momentum along the $x$ direction. The computation proceeds in much the same way as before
and we obtain the following result for the conductivity, valid (for any $\om$) to linear order in $\gamma$,
\beq
\sigma = \frac{1}{g_4^2} \left(1+\frac{16 \gamma}{L^2} \right), \;\;\;\;\; D \sim \frac{3}{4 \pi T} \left(1+\frac{6\gamma}{L^2} \right),
\eeq
which for $\gamma\rightarrow 0$ is consistent with existing results \cite{Herzog:2002fn}. We have also exhibited the result for the diffusion
constant obtained using the membrane current prescription (\ref{eq:generalD}). We observe that the correction to the conductivity is
independent of  temperature as expected, and is also independent of  $\om$ to linear order in the perturbation. 

\subsection{Corrections to conductivity in 6 dimensions}
Repeating the above calculation in $d=6$ dimensions, using the background (\ref{adsbh}) 
corresponding for example to the near-horizon geometry of a stack of black M5-branes, we require the following nonzero components of the Weyl tensor,
\beqn
C_{0i0j} &=& \frac{f \alpha^4 L^2 u}{2} \delta_{ij}, \hspace{1cm} C_{0u0u} = \frac{5 \alpha^2 L^2}{2}, \nonumber \\
C_{iuju} &=& -\frac{\alpha^2 L^2}{2f} \delta_{ij}, \hspace{1.3cm} C_{ijkl} = \frac{\alpha^4 L^2 u}{4}\delta_{ik}\delta_{jl},
\eeqn
where $u=(r_0/r)^2$ and $\al=4\pi T/3$ with $f(u)=1-u^3$. Aligning the gauge field to propagate along one of the five spatial
directions, we are again able to solve the equations for the transverse retarded correlator and can read off the conductivity,
\beq
\sigma = \frac{\alpha^3 R^3}{g_7^2} \left(1+\frac{4 \gamma}{L^2} \right), \;\;\;\;\; D \sim \frac{3}{8\pi T} \left(1+\frac{48 \gamma}{L^2} \right).
\eeq
Note that the leading dependence of $\si$, when written in terms of M5 worldvolume parameters, is $T^3 N^3$ as expected for
the scaling of transport parameters in this case \cite{Herzog:2002fn}. We have again quoted the result for the diffusion constant
from (\ref{eq:generalD}) for comparison.

\section{Discussion}

Given the set of $\gamma$-dependent corrections to $\si$ and $D$ discussed in the preceding section, 
we will conclude by discussing some issues that go beyond the generic picture of effective field theory used in
the paper. In particular, we will address possible constraints on $\gamma$ that arise from considerations of bulk and boundary causality.

\begin{itemize}
\item {\bf Causality constraints}:
Given that the Weyl coupling arises in a more complete theory from a locally Lorentz-invariant UV completion, we may ask whether 
causality places interesting constraints on $\gamma$, and more generally on the structure of the tensor $X^{\mu\nu\rh\si}$. This issue,
particularly within the context of QED in curved space, has been studied in some detail \cite{dh,hs,dntv}. However, while in flat space
the relation between superluminal propagation and causality violation may in fact lead to interesting constraints on the effective
field theory expansion \cite{aadnr}, the issue appears to be more subtle in curved space. The curvature coupling $CFF$ is birefringent,
and so it will in general ensure, regardless of the sign of $\ga$, that one polarization is superluminal as observed in \cite{dh}. 
Taking an eikonal limit  for a solution with polarization vector $a_\mu$ and momentum $q^\nu$, i.e.  $\om,\vec{q}\rightarrow\infty$, we find that
it satisfies $g^{\mu\nu}_{\rm eff} q_\mu q_\nu=0$ and hence propagates according to  the null cone of an effective metric 
$g^{\mu\rh}_{\rm eff}=g^{\mu\rh} - 8\gamma C^{\mu\nu\rh\si}a_\nu a_\si$. For example, the phase velocity for an $x$-polarized mode propagating in the 
$z$-direction at fixed radius  takes the form
\be
 v_{\rm ph}^2(u) = f(u) \left(\frac{1- 8\gamma u^2/L^2}{1+8\gamma u^2/L^2}\right), \label{vph}
\ee
which allows superluminal propagation for $\gamma < -L^2/16$. However, even though it was obtained in an eikonal limit, if we treat this system
as an effective field theory, (\ref{vph}) refers to frequencies which
are necessarily small relative to the effective theory cutoff, i.e. $\om \sqrt{\gamma} \ll 1$. In curved space, superluminal modes present in
this regime do not directly reflect a violation of causality. For that we need to consider the wavefront velocity 
$v_{\rm wf} = v_{\rm ph}(\om\rightarrow \infty) =  1$ which is not accessible within the effective theory \cite{hs}, or more generally 
explore the region of support for causal Green's functions \cite{dntv}.

It is nonetheless interesting to take the Weyl-corrected Maxwell action more literally, as the AdS/CFT correspondence in principle allows us to
consider any generic action for the field $A_\mu$ which couples to the conserved U(1) current. Taking this viewpoint,
we can ask whether the fixed boundary causal structure can be used to infer additional constraints, following the argument of Brigante et al.
\cite{myers}. In particular, 
the equation for $A_x$ given in (\ref{eom}) can be recast as a radial 
Schr\"odinger equation with a potential of the form $V(\tilde{u}) = (\vec{q}/2\al)^2 v_{\rm ph}^2(\tilde{u}) + V_1(\tilde{u})$ where $d\tilde{u}/du=1/(f\sqrt{u})$ is 
a monotonic change of coordinates. It follows that for large $q^2$ the potential is dominated by the $v_{\rm ph}^2q^2$ term, apart from
a small region near the boundary ($\tilde{u}\rightarrow 0$) where $V_1(\tilde{u})\sim 1/\tilde{u}^2$. Consequently, from the form of  
$v_{\rm ph}^2(u)$ in (\ref{vph}), we find that for $\gamma < -L^2/16$ and $q^2$ sufficiently large the potential develops a local minimum and 
a metastable bound state is possible. This reflects the presence of
a long-lived CFT mode which in this regime can be used for superluminal transport, as discussed  in \cite{myers}. Consequently, given the
fixed boundary causal structure, the presence of this mode links superluminal bulk velocities to a violation of (boundary) causality. The ensuing
causality constraint $\gamma > -L^2/16$ leads to the bounds on $\si$ and $D$ shown in (\ref{limits}) for the CFT dual to the bulk theory with
this specific Weyl coupling. However, it is important to emphasize that, for $\gamma$ near the lower bound, we need to 
take $q^2 \gg 1/\sqrt{\gamma}$ which is beyond the effective field theory cutoff, and so this conclusion holds provided we 
ignore possible higher order bulk corrections. It is nonetheless intriguing that, within the regime of validity of the 1-loop calculation of \cite{dh},
the correction to $\gamma$ (\ref{1loop}) from a massive threshold is actually consistent with this causality constraint (now in $d=3$), even 
though these terms apparently allow for superluminal propagation in other backgrounds. Indeed, it is known from more subtle analyses 
that  in this case the wavefront velocity is  not superluminal and thus causality is not in jeopardy \cite{hs,dntv}. 

Given this novel viewpoint on bulk causality, we note that the lower bound on $\gamma$ discussed above may actually be supplemented
with an upper bound that arises from noting that certain modes may become ghost-like near the horizon. This is already apparent
in (\ref{vph}) as the phase velocity for this mode can become negative for $\gamma > L^2/8$. Moreover, the
bulk kinetic functions for $A_x$ and $A_r$ are determined by $X^{x0x0}$ and $X^{r0r0}$ respectively, and evaluating the components we 
see that these modes can become ghost-like near the horizon unless $-\frac{L^2}{8} < \gamma < \frac{L^2}{24}$.  While these constraints
necessarily appear somewhat gauge-dependent, they are indicative  of possible problems. Indeed this somewhat weaker lower
bound on $\gamma$ is reflected in the dual theory by the vanishing of $\si$ and $D$, while they are formally negative for lower values of
$\gamma$. Putting these pieces together, we obtain the following range,
\be
 -\frac{L^2}{16} < \gamma < \frac{L^2}{24},
\ee
as the strongest constraint on the Weyl-corrected Maxwell theory that can be inferred from our simple causality arguments. 
As discussed above, the lower limit here may be on the most solid footing and suggests the intriguing lower bounds on $\si$ and $D$ discussed in Section~1.
 It would clearly be interesting to explore these constraints
in greater depth.\footnote{Note added (v2, 10 July 2009): The constraint $\gamma > -L^2/16$ was also obtained in \cite{hm} using different considerations, namely 
requiring positivity of
energy expectation values. The relation between these constraints has recently been discussed in \cite{hofman09}.}

\item {\bf Other saddle points:} Beyond the causality constraints on $\gamma$, there may be others that require closer inspection. Indeed, we have implicitly
assumed that the background solution with $F=0$ remains the dominant saddle point in the ensemble. However, it is well-known that
other solutions do exist, e.g. if we work in the grand canonical ensemble and turn on a chemical potential, the dual bulk geometry is
a charged black hole. It would be interesting to know if, on increasing $\gamma$, other solutions may be possible and whether any
of these solutions might become the dominant saddle point.  
\end{itemize}

\acknowledgments
\noindent
We thank Pavel Kovtun for numerous helpful discussions and comments on the manuscript.
This work was supported in part by NSERC of Canada.


\end{document}